\documentclass{PoS}

\title{Recent results on event-by-event fluctuations in ALICE at the LHC}

\ShortTitle{Fluctuations results in ALICE}

\author{\speaker{Nirbhay Behera} for the ALICE collaboration%
         \\
        IIT Bombay\\
        E-mail: \email{nbehera@cern.ch}}

\abstract{Non-statistical event-by-event fluctuations in relativistic heavy-ion
collisions have been proposed as a probe of the phase transition of hadronic matter to a deconfined phase of quarks and gluons, the so-called Quark-Gluon Plasma. In a thermodynamical picture of the strongly
interacting system formed in heavy-ion collisions, the dynamical
fluctuations of net-charge, fluctuations of the mean transverse momentum, mean
multiplicity and balance functions are related to the fundamental
properties of the system, hence they may reveal information about the
QCD phase transition. In this article, recent
results on event-by-event measurements of net-charge fluctuations, the measurement of the balance function and mean transverse momentum fluctuations are discussed.
}

\FullConference{9th International Workshop on Critical Point and Onset of Deconfinement - CPOD2014,\\
		17-21 November 2014\\
		ZiF (Center of Interdisciplinary Research), University of Bielefeld, Germany}

\begin{document}

%------------------INTRODUCTION------------
\section{Introduction}
Heavy-ion collision experiments at the Relativistic Heavy-Ion Collision (RHIC) and the Large Hadron Collider (LHC) aim to study the phase transition of hadronic matter to the Quark-Gluon Plasma (QGP) phase and to characterize its properties. It is proposed that the study of event-by-event fluctuations and correlations are sensitive tools to look for evidence of a possible phase transition and its thermodynamic properties \cite{1,2,3}. The measurement of fluctuations of particle multiplicities and their ratios, fluctuations of conserved charge (e.g. net-charge, net-proton), two-particle correlations, mean transverse momentum fluctuations are common tools used in heavy-ion collision experiments to study the correlations and fluctuations originating from the system produced.
\par
In this article, we discuss the recent results on event-by-event fluctuations measured at mid-rapidity with the ALICE \cite{3a} experiment at the LHC. The Time Projection Chamber (TPC) is used as the main tracking detector for these analyses and the Inner Tracking System (ITS) is used for vertex determination. The particle multiplicity per event measured with a set of scintillator arrays (VZERO) is used to determine the collision centrality. We have also compared our measurements with results obtained at lower collision energies from the RHIC. The article is organized as follows. First we discuss the results of net-charge fluctuations in Section 2. In Section 3, the results on the balance function are presented. In Section 4, the results of mean transverse momentum fluctuations are discussed. Finally in section 5, we summarize the findings and provide an outlook.

%-------------Net-charge ------------------------
\section{Net-charge fluctuations}
It has been proposed that there is a significant reduction of net-charge fluctuations from the QGP medium to the final state hadronic medium due to the entropy conservation \cite{1,2,4}. This can be argued in a very simple way as follows. In the QGP medium, quarks are the charge carriers, which have fractional electric charge, whereas the hadron gas has integral (unit) charge. So the variance of fluctuations of net-charge ($\delta Q^2$) measured in the QGP medium will be substantially smaller than in the hadronic medium assuming that quark-quark correlation is negligible and that hadronization of gluons produces pairs of positive and negative charges hence doesn't contribute to the final state charge fluctuations. Thus the measurement of net-charge fluctuations can be regarded as a sensitive indicator of the formation of QGP in heavy-ion collisions \cite{3}. However, there may be fluctuations of system size, which may affect the net-charge fluctuations. These volume fluctuations can be cancelled out by measuring the fluctuations of the ratio of positive ($N_{+}$) to negative ($N_{-}$) charges \cite{3}. Theoretically, the ratio ($R = N_{+}/N_{-}$) can be related to the D measure, which links the charged fluctuation per unit entropy as follows \cite{5,6}.

\begin{equation}
{\rm D} = \langle N_{ch} \rangle \langle \delta R^2 \rangle \approx 4 \frac{\langle \delta Q^2 \rangle}{\langle N_{ch} \rangle} ~,
\end{equation}

where $\langle \delta R^2 \rangle = \langle R^2 \rangle - \langle R \rangle^2$, $\delta Q^2$ is the variance of fluctuations of net-charge ($Q= N_{+} - N_{-}$) and $N_{ch}$ is the total charge ($N_{ch} = N_{+} - N_{-}$) measured at the mid-rapidity. The value of D has been estimated from several theoretical works both for QGP and hadron gas (HG). According to Ref.\cite{1}, the value of D in a HG is approximately 4 times larger than in a QGP. Taking quark-quark interactions, lattice QCD calculations predict different values of D for different media. For an uncorrelated pion gas, D = 4 and after taking resonances into account, the value of D decreases to 3, for the QGP phase, D $\simeq 1.0 - 1.5$ \cite{4a}. The uncertainty in the D value for the QGP phase comes from the formulation of relating the entropy to the number of charged particles in the final state. Experimentally, D is estimated by measuring the dynamical fluctuations of the net-charge, $\nu_{(+-,dyn)}$, as follows \cite{6,4a}.

\begin{equation}
\langle N_{ch} \rangle \nu_{(+-,dyn)} \approx {\rm D} - 4 ~,
\end{equation}

where $\nu_{(+-,dyn)}$ is given as

\begin{equation} 
\nu_{(+-,dyn)} = \frac{ \langle N_{+}(N_{+} - 1) \rangle}{ {\langle N_{+} \rangle}^2 } + \frac{ \langle N_{-}(N_{-} - 1) \rangle }{ {\langle N_{-} \rangle}^2 } - 2 \frac{ \langle N_{-}N_{+}\rangle }{ \langle N_{-} \rangle  \langle N_{+} \rangle} ~.
\end{equation}

$\nu_{(+-,dyn)}$ measures the relative correlation strength of particle pairs and has been found robust against the detector efficiency \cite{4b}. Before correlating the measured $\nu_{(+-,dyn)}$, it needs to be corrected for global charge conservation (GCC) and finite net-charge effect \cite{4b}. So the corrected $\nu_{(+-,dyn)}$ will be, 

\begin{equation}
\nu_{(+-,dyn)}^{corr} = \nu_{(+-,dyn)} + \frac{4}{\langle N_{total}\rangle} ~.
\end{equation}
$\langle N_{total}\rangle$ is the number of total charged particles produced in the whole phase space which is estimated from experimental data. Therefore, taking this GCC correction, D can be written as,
\begin{equation}
{\rm D} = \langle N_{ch} \rangle \nu_{(+-,dyn)}^{corr} + 4 ~.
\end{equation}

\begin{figure}[h]
\begin{center}
\begin{minipage}{16pc}
\begin{center}
\includegraphics[width=16pc]{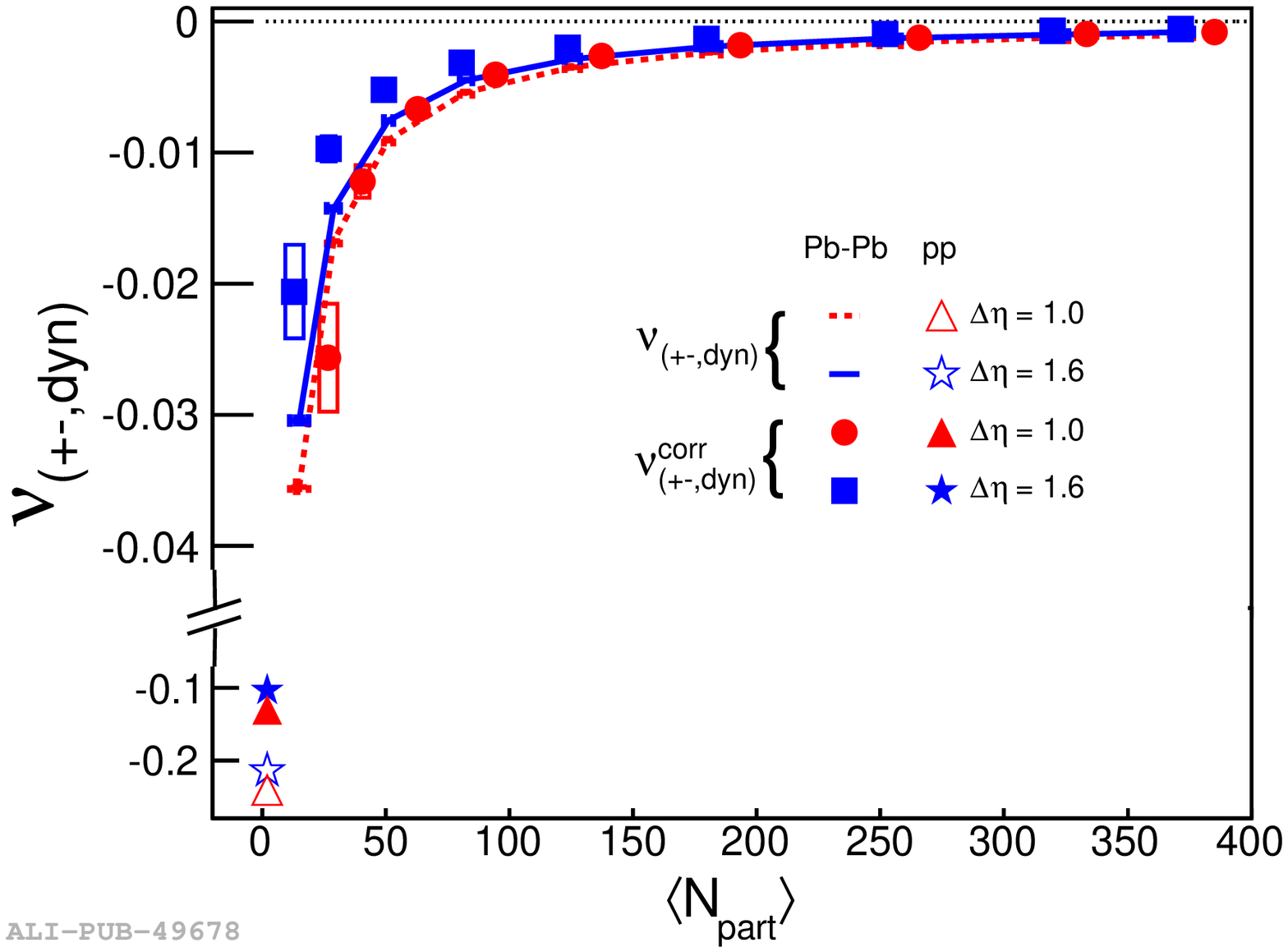}
(a)
\end{center}
\end{minipage}\hspace{1pc}%
\begin{minipage}{16pc}
\begin{center}
\includegraphics[width=16pc]{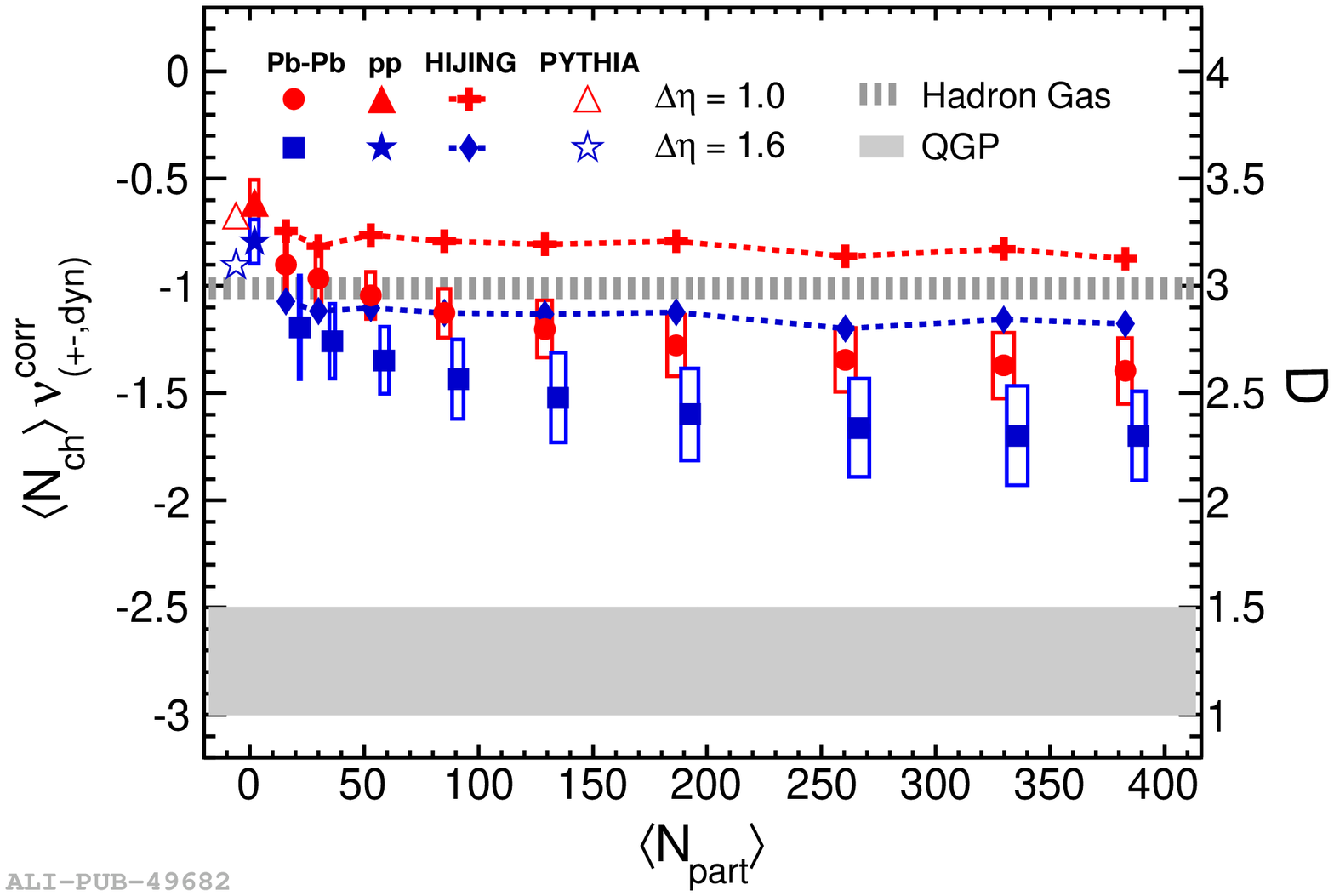}
(b)
\end{center}
\end{minipage} 
\caption{\label{fig1} (a): $\nu_{(+-,dyn)}$ and $\nu_{(+-,dyn)}^{corr} $ as a function of $\langle N_{part} \rangle$. (b): $\langle N_{ch} \rangle \nu_{(+-,dyn)}^{corr} $ as a function of $\langle N_{part} \rangle$. The dashed line and solid band represents the value of D for hadron gas and QGP, respectively. Data are compared with simulation results obtained from HIJING \cite{hijing} and PYTHIA \cite{pythia}. Data are obtained for Pb$-$Pb collisions at $\sqrt{s_{NN}} =$ 2.76 TeV and for pp collisions at $\sqrt{s} =$2.76 TeV. The statistical and systematic error bars are represented by vertical lines and boxes, respectively. Figures are taken from Ref. \cite{6}}
\end{center}
\vskip -12pt
\end{figure}

We have measured net-charge fluctuations by calculating $\nu_{(+-,dyn)}$  and D as a function of collision centrality for minimum bias events in pp and Pb$-$Pb collisions at $\sqrt{s_{NN}} =$ 2.76 TeV. The dynamical net-charge fluctuations, $\nu_{(+-,dyn)}$ and $\nu_{(+-,dyn)}^{corr}$ for charged particles as a function of $\langle N_{part} \rangle$ are shown in Figure \ref{fig1} (a). The results for $\langle N_{ch} \rangle \nu_{(+-,dyn)}^{corr} $ as a function of $\langle N_{part} \rangle$ for pp and  Pb$-$Pb collisions are compared to results obtained from models, like HIJING \cite{hijing} and PYTHIA \cite{pythia} event generators, which are shown in Figure \ref{fig1} (b).
%------------------------------------------------------------------
%----------------------------------------
%\begin{figure}[!h]
%\begin{center}$
%\begin{array}{cc}
%\includegraphics[width=2.7in]{2013-Jun-18-figure1.eps}
%\includegraphics[width=2.7in]{2013-Jun-18-figure2.eps}
%\end{array}$
%\end{center}  
%\caption{Left: $\nu_{(+-,dyn)}$ and $\nu_{(+-,dyn)}^{corr} $ as a function of $\langle N_{part} \rangle$. Right: $\langle N_{ch} \rangle \nu_{(+-,dyn)}^{corr} $ as a function of $\langle N_{part} \rangle$. The dashed line and solid band represents the value of D for hadron gas and QGP, respectively. Data are compared with simulation results obtained from HIJING and PYTHIA. The statistical and systematic error bars are represented by vertical lines and boxes, respectively. Figures are taken from Ref. \cite{6}}
%\label{fig1}
%\end{figure}

\begin{figure}[h]
\begin{center}
\begin{minipage}{16pc}
\begin{center}
\includegraphics[width=16pc]{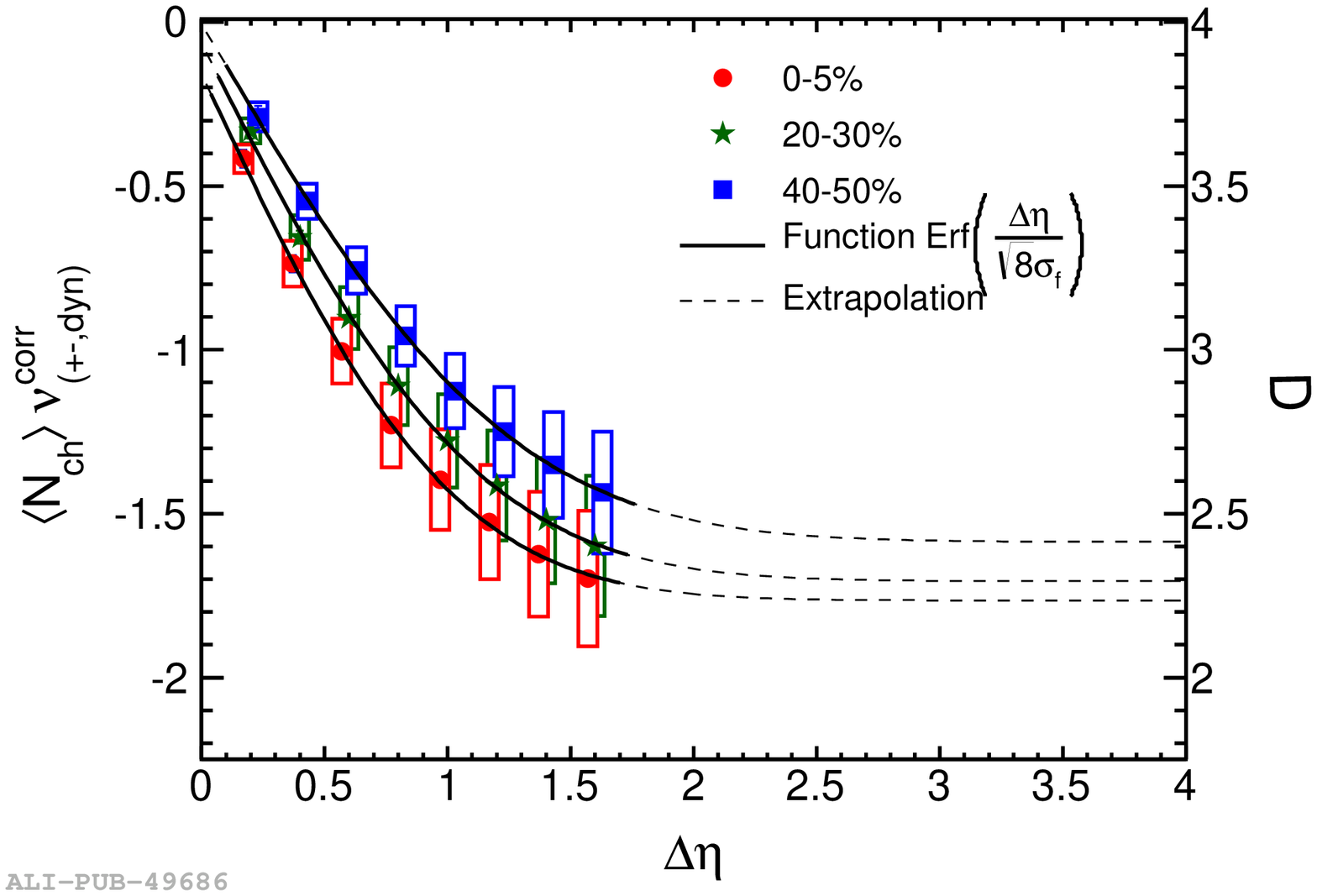}
(a)
\end{center}
\end{minipage}\hspace{1pc}%
\begin{minipage}{16pc}
\begin{center}
\includegraphics[width=16pc]{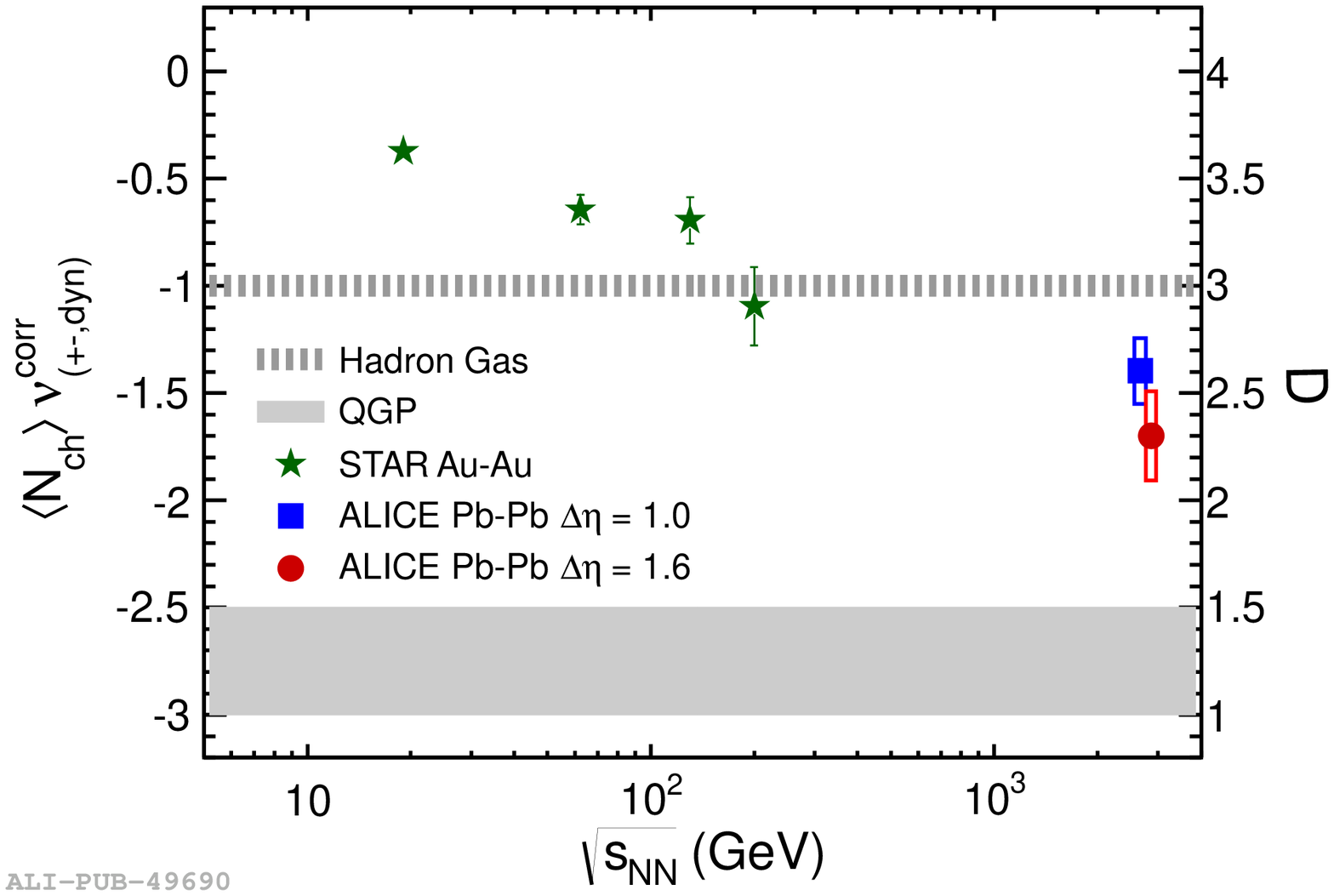}
(b)
\end{center}
\end{minipage} 
\caption{\label{fig2} (a): $\langle N_{ch} \rangle\nu_{(+-,dyn)}^{corr}$ (in left y-axis) and the D measure (in right y-axis) as a function of $\Delta \eta$ fitted with error function (solid line) for three collision centralities. The dashed lines are extrapolated to higher $\Delta \eta$ values. (b): The energy dependence of $\langle N_{ch} \rangle\nu_{(+-,dyn)}^{corr}$ and D for the top central collisions. The STAR results are taken from Ref. \cite{4}.  Systematic errors are represented by boxes. Figures are taken from Ref.\cite{6}}
\end{center}
\vskip -12pt
\end{figure}

The $\nu_{(+-,dyn)}^{corr}$  is measured taking two pseudorapidity intervals, i.e. $\Delta \eta =$ 1.0 (-0.5 $< \eta <$ 0.5) and $\Delta \eta =$ 1.6 (-0.8 $< \eta <$ 0.8). It can be seen from Figure \ref{fig1} (b) that HIJING is unable to explain the Pb$-$Pb data. For a particular centrality range, the D value shows a decreasing trend with increasing pseudorapidity window. This suggests a strong dependence of dynamical charge fluctuations on the pseudorapidity window. In Ref. \cite{7,8}, this behavior is explained by considering the concept of diffusion in rapidity space. It is shown that there is a dissipation of the fluctuations originating from the initial stage of the collision during the evolution of the system. This nature of dissipation of the primordial fluctuation signal can be described by an error function of form erf$(\Delta \eta / \sqrt{8\sigma_f})$. In the error function, $\sigma_f$ represents the diffusion at freeze-out. We fit the data as a function of $\Delta \eta$ with that error function, which is shown in Figure \ref{fig2} (a). From the fit, it can be seen that there is a decrease in the magnitude of fluctuations with increase in $\Delta \eta$ and it tends to saturate at $\Delta \eta = $ 2.3. In addition, the energy dependence of $\langle N_{ch} \rangle \nu_{(+-,dyn)}^{corr} $ and D measure as a function of $\sqrt{s_{NN}}$ is given in Figure \ref{fig2} (b). A monotonic decrease of the D value as a function of collision energy is observed. More importantly, the ALICE measurement for $\Delta \eta =$ 1.6 is closer to the QGP values than the other measurements at lower energies for $\Delta \eta =$ 1.0.

%\begin{figure}[!h]
%\begin{center}$
%\begin{array}{cc}
%\includegraphics[width=2.7in]{2013-Jun-18-figure4.eps} &
%\includegraphics[width=2.7in]{2013-Jun-18-figure3.eps}
%\end{array}$
%\end{center}  
%\caption{Left: The energy dependence of $\langle N_{ch} \rangle\nu_{(+-,dyn)}^{corr}$ and D for the top central collisions. The STAR result is taken from Ref. \cite{4}. Right: $\langle N_{ch} \rangle\nu_{(+-,dyn)}^{corr}$ (in left y-axis) and D measure (in right y-axis) as a function of $\Delta \eta$ fitted with error function (solid line) for three collision centralities. The dashed lines are extrapolated to higher $\Delta \eta$ values. Systematic errors are represented by boxes. Figures are taken from \cite{6}}
%\label{fig2}
%\end{figure}
 
 %-------------Balance function---------------------- 
\section{Balance function}
The measurement of charge dependent two-particle correlations via the so-called balance function  is argued to probe the charge creation time and the collective motion of the system \cite{8a,9}. The general definition of the balance function is given as follows \cite{10}.

\begin{equation}
B_{ab}(P_2, P_1) = \frac{1}{2} \left[ C_{ab}(P_2,P_1) + C_{ba}(P_2,P_1) - C_{bb}(P_2,P_1) - C_{aa}(P_2,P_1) \right] ~,
\end{equation}
where $C_{ab}(P_2,P_1) = N_{ab}(P_2,P_1)/N_b(P_1)$ is the distribution of the pairs of particles, of type $a$ and $b$, are in momentum space $P_1$ and $P_2$, respectively, normalized to the numbers of particles $b$. Here particles $a$ and $b$ refer to all positive and negative particles, respectively. If the correlation is measured in terms of pseudorapidity difference $\Delta \eta = |\eta_b - \eta_a |$, the balance function is given as,

\begin{equation}
B_{+-}(\Delta \eta) = \frac{1}{2} \left[ C_{+-}(\Delta \eta) + C_{-+}(\Delta \eta) - C_{--}(\Delta \eta) - C_{++}(\Delta \eta) \right] ~.
\label{Beta}
\end{equation}

\begin{figure}[h]
\begin{center}
\begin{minipage}{16pc}
\begin{center}
\includegraphics[width=16pc]{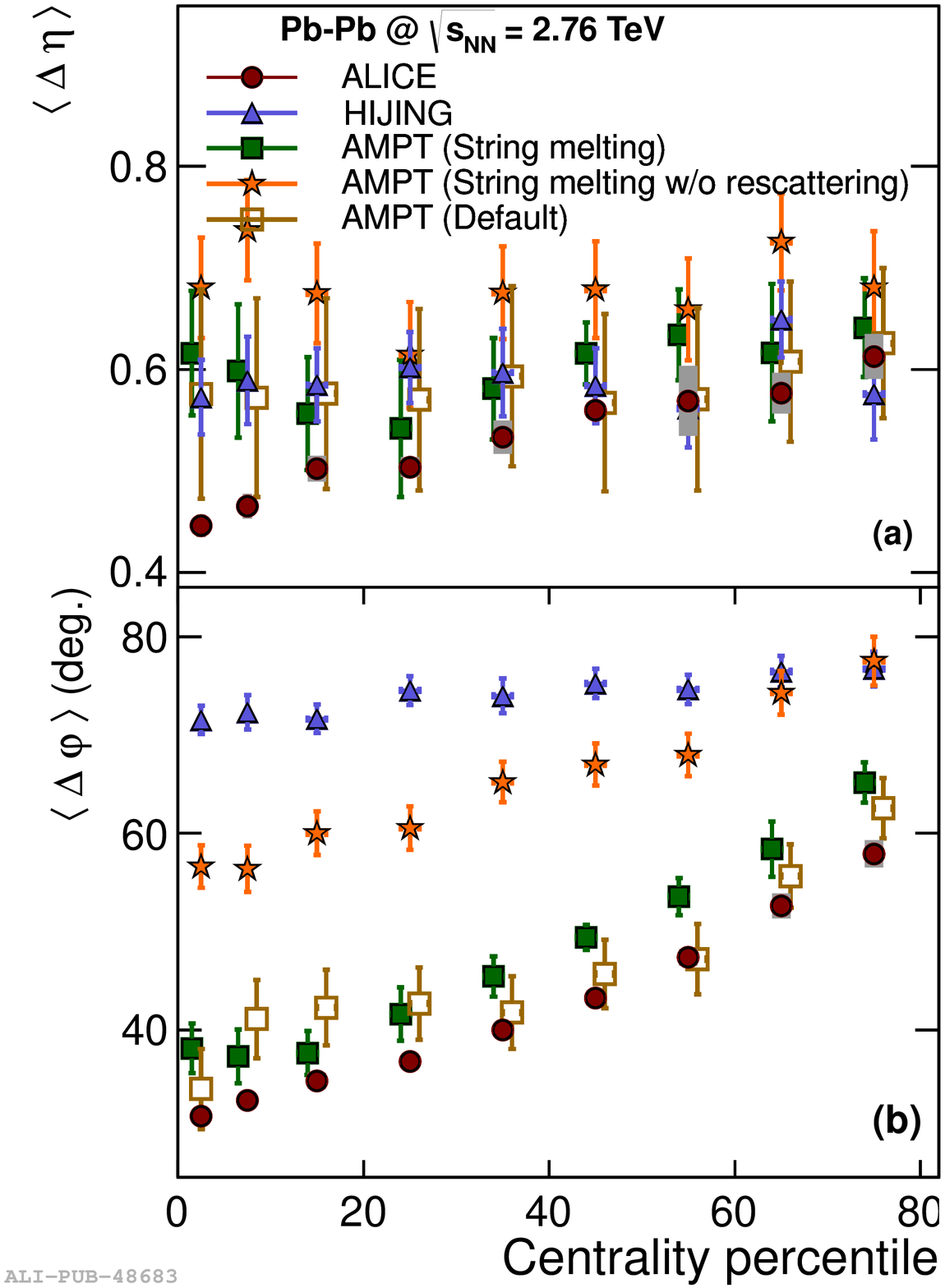}
(a)
\end{center}
\end{minipage}\hspace{1pc}%
\begin{minipage}{16pc}
\begin{center}
\includegraphics[width=16pc]{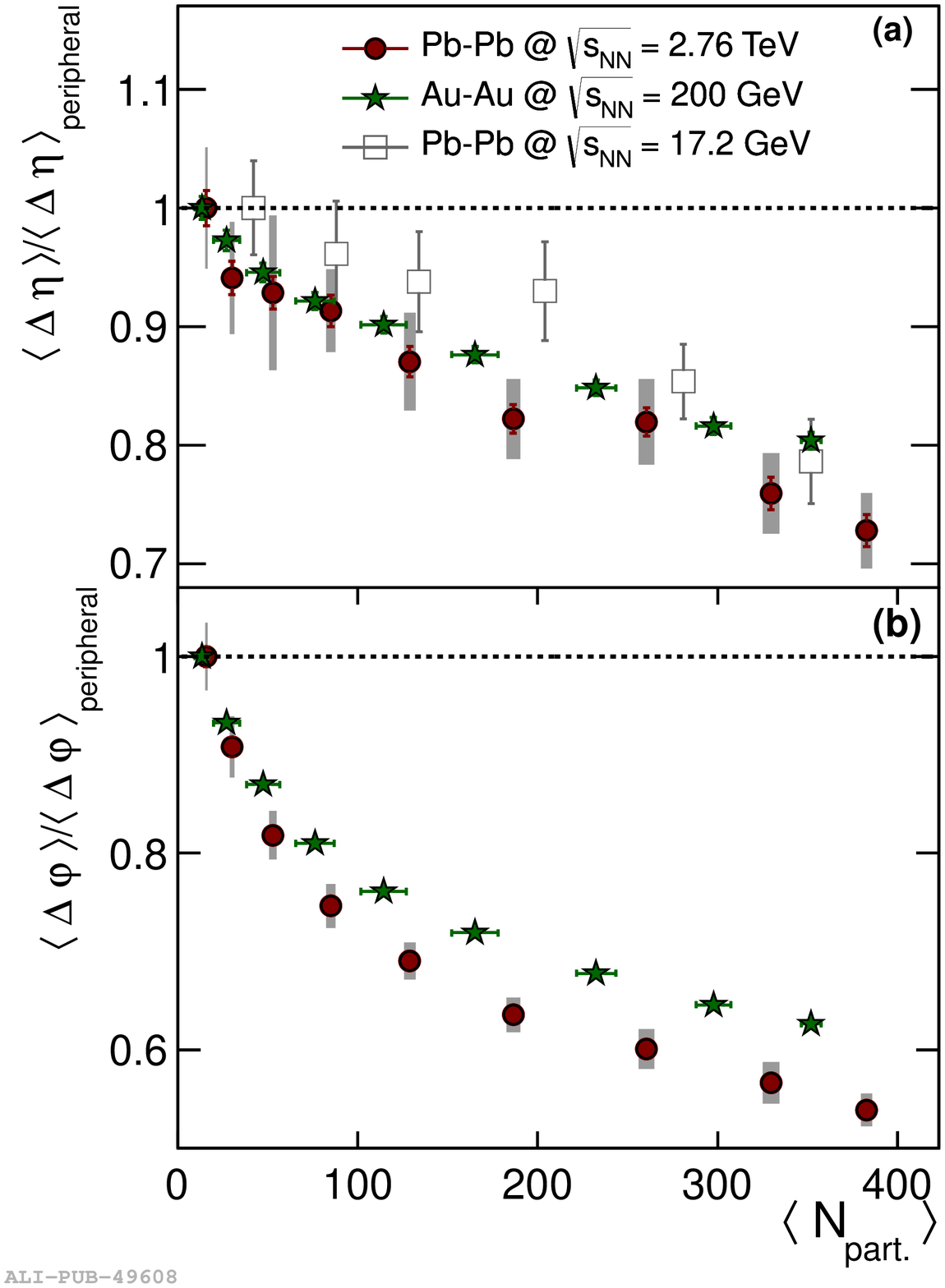}
(b)
\end{center}
\end{minipage} 
\caption{\label{balnceFuntion} (a): Centrality dependence of the width of the balance function $\langle \Delta \eta \rangle$ and $\langle \Delta \phi \rangle$, studied in terms of relative pseudorapidity and relative azimuthal angle for charged particles at mid-rapidity. The data are compared with predictions from HIJING and different configurations of AMPT. (b): The relative decrease of width of the balance function in the relative pseudorapidity and azimuthal angle as a function of centrality. ALICE data are compared with top SPS and RHIC energies. Figures are taken from Ref.\cite{10}}
\end{center}
\vskip -12pt
\end{figure}

Each term of Eq. \ref{Beta} is corrected for detector acceptance and tracking efficiency \cite{10}. Similarly, Eq. \ref{Beta} can be written for azimuthal angle ($\Delta \phi = | \phi_b - \phi_a |$). ALICE has measured the widths of balance function which are characterized by $\langle \Delta \eta \rangle$ and $\langle \Delta \phi \rangle$ studied in pseudorapidity and azimuthal angle, respectively. For example, $\langle \Delta \eta \rangle$ is calculated as follows.

\begin{equation}
\langle \Delta \eta \rangle = \frac{ \sum_{i=1}^{k} \left[ B_{+-}(\Delta \eta_{i}) .\Delta \eta_{i} \right]} { \sum_{i=1}^{k} \left[ B_{+-}(\Delta \eta_{i}) \right] } ~,
\end{equation} 
where $B_{+-}(\Delta \eta_{i})$ is the balance function value for each bin $\Delta \eta_{i}$ with the sum running over all bins $k$. The ALICE results of balance function in Pb$-$Pb collisions at $\sqrt{s_{NN}} =$ 2.76 TeV as a function of centrality percentile are shown in Figure \ref{balnceFuntion} (a). The data are compared with the Monte Carlo simulation results of HIJING \cite{hijing} and AMPT \cite{ampt}. It is found that the widths of the balance function measured for $\langle \Delta \eta \rangle$ and $\langle \Delta \phi \rangle$ decrease while moving from peripheral to central collisions. The results could not be reproduced by HIJING, whereas AMPT tuned to describe the $v_2$ values reported by ALICE \cite{10a} can qualitatively reproduce the centrality dependence of $\langle \Delta \eta \rangle$ but fails to describe the $\langle \Delta \phi \rangle$ values. The relative decrease of width of the balance function in the relative pseudorapdity and azimuthal angle from peripheral to central collisions as a function of $\langle N_{part} \rangle$ is shown in Figure \ref{balnceFuntion} (b). The ALICE points are compared with the results obtained from highest SPS and RHIC energies. The correlation in relative pseudorapidity ($\langle \Delta \eta \rangle / \langle \Delta \eta \rangle_{peripheral}$) and in relative azimuthal angle ($\langle \Delta \phi \rangle / \langle \Delta \phi \rangle_{peripheral}$) shows a decrease of $\approx (9.5\pm 2.0(stat)\pm 2.5(sys))$ $\%$ and $\approx (14.0\pm 1.3(stat)\pm 1.9(sys)$)$\%$ respectively, for the most central collisions compared to RHIC results. The latter can be understood as additional increase in radial flow between central and peripheral collisions at the LHC compared to RHIC. 

%-----------------------------------------------------
%-----------------------------------------------------
%\begin{figure}[!h]
%\begin{center}$
%\begin{array}{cc}
%\includegraphics[width=2.7in]{2013-May-10-figure3a_CorrEM.eps} &
%\includegraphics[width=2.7in]{2013-Jun-17-figure6_CorrEM.eps}
%\end{array}$
%\end{center}  
%\caption{Left: Centrality dependence of the width of the balance functions $\langle \Delta \eta \rangle$ and $\langle \Delta \phi \rangle$, studied in terms of relative pseudorapidity and relative azimuthal angle. The data are compared with predictions from HIJING, AMPT $default$ and AMPT $string~melting$. Right: The relative decrease of width of the balance functions in the relative pseudorapidity and azimuthal angle as a function of centrality. ALICE data are compared with top SPS and RHIC energies. Figures are taken from Ref.\cite{10}}
%\label{balnceFuntion}
%\end{figure}

%---------------mean pT fluctuations------------------
\section{Mean $p_{T}$ fluctuations}

The measurement of event-by-event fluctuations of mean $p_{\rm T}$ ($\langle p_{\rm T} \rangle$) is proposed as an excellent tool to characterize the thermodynamics and collectivity of the system produced in heavy-ion collisions \cite{1,2,3}. Mean  $p_{\rm T}$ fluctuations measured in pp collisions also serve as a model independent baseline to figure out the non-trivial fluctuations in A-A collisions. The event-by-event mean transverse momentum is measured by the mean value of $M_{\rm EbE}(p_{\rm T})_{k}$ of the transverse momenta $p_{\rm T,i}$ of the $N_{acc},k$ accepted charged particles in event $k$ as follows.
\begin{equation}
M_{\rm EbE}(p_{\rm T})_{k} = \frac{1}{N_{acc,k}} \sum_{i=1}^{N_{acc},k} p_{\rm T,i}.
\end{equation}

The event-by-event fluctuations of $M_{\rm EbE}(p_{\rm T})_{k}$ have both dynamical and statistical contributions. The dynamical component is measured by the two-particle correlator $C_m = \langle \Delta p_{\rm T,i}, \Delta p_{\rm T,j} \rangle $ and is defined as follows \cite{11}.

\begin{equation}
C_{m} = \frac{1}{\sum_{k=1}^{n_{ev,m}} N_{k}^{pairs}} \cdot \sum_{k=1}^{n_{ev,m}} \sum_{i=1}^{N_{acc,k}} \sum_{j=i+1}^{N_{acc,k}}  \left(  p_{\rm T,i} - M(p_{\rm T})_m \right) \cdot \left( p_{\rm T,j} - M(p_{\rm T})_m \right)~,
\end{equation}
where $n_{ev,m}$ is the number of events in a given multiplicity class $m$, $N_{k}^{pairs}$ is the number of pairs constructed out of $N_{k}$ number of particles in an event and equal to 0.5$N_{k}(N_{k}-1)$. $M(p_{\rm T})_{m}$ is the average $p_{\rm T}$ of all tracks in all events of multiplicity class m, which is defined as follows.

\begin{equation}
M(p_{\rm T})_{m} = \frac{1}{\sum_{k=1}^{n_{ev,m}} N_{acc,k}} \sum_{k=1}^{n_{ev,m}} N_{acc,k} \cdot M_{EbE}(p_{\rm T})_{k} ~.
\end{equation}

\begin{figure}[h]
\begin{center}
\begin{minipage}{12pc}
\begin{center}
\includegraphics[width=12pc]{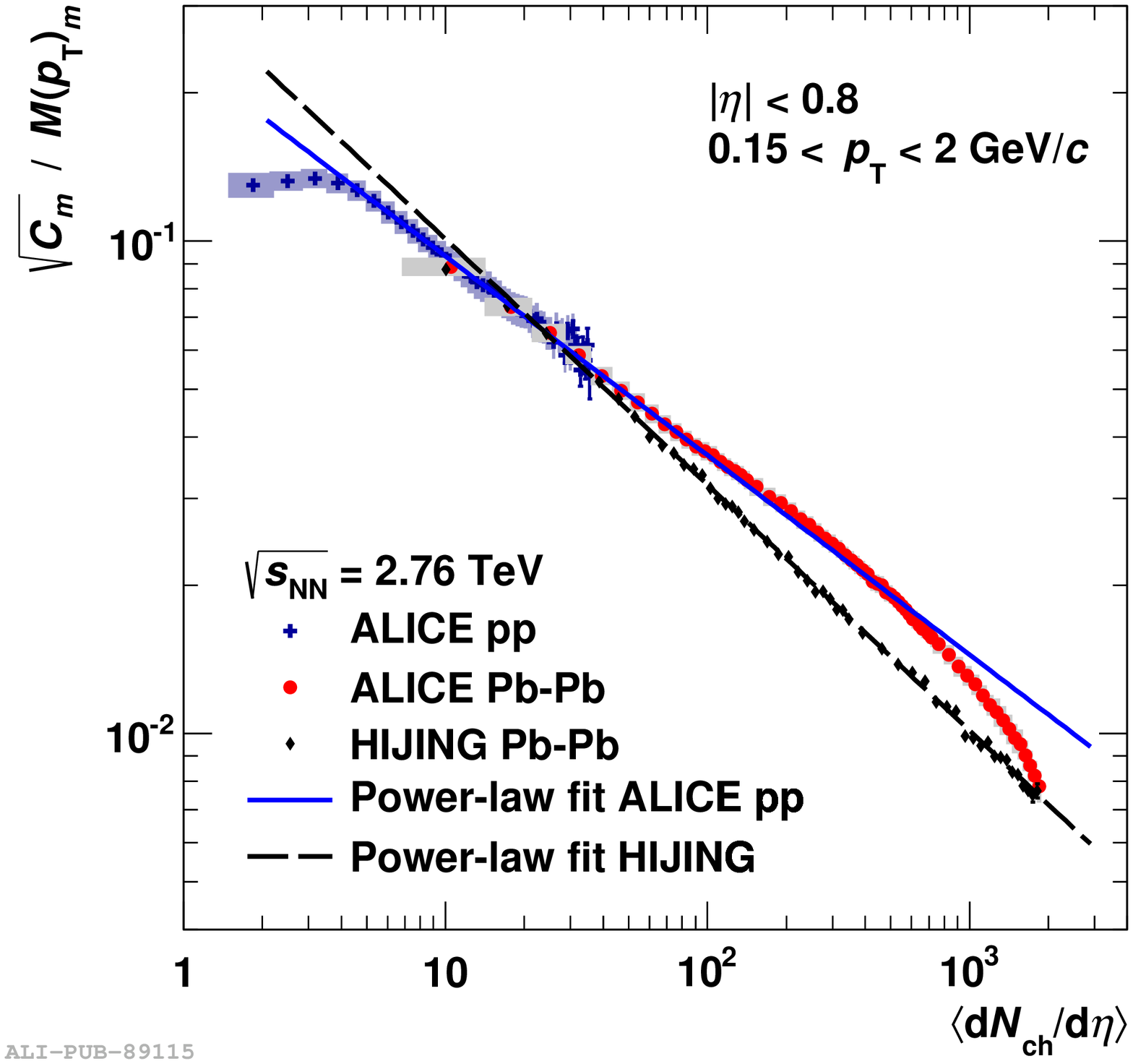}
(a)
\end{center}
\end{minipage}\hspace{1pc}%
\begin{minipage}{19pc}
\begin{center}
\includegraphics[width=20pc]{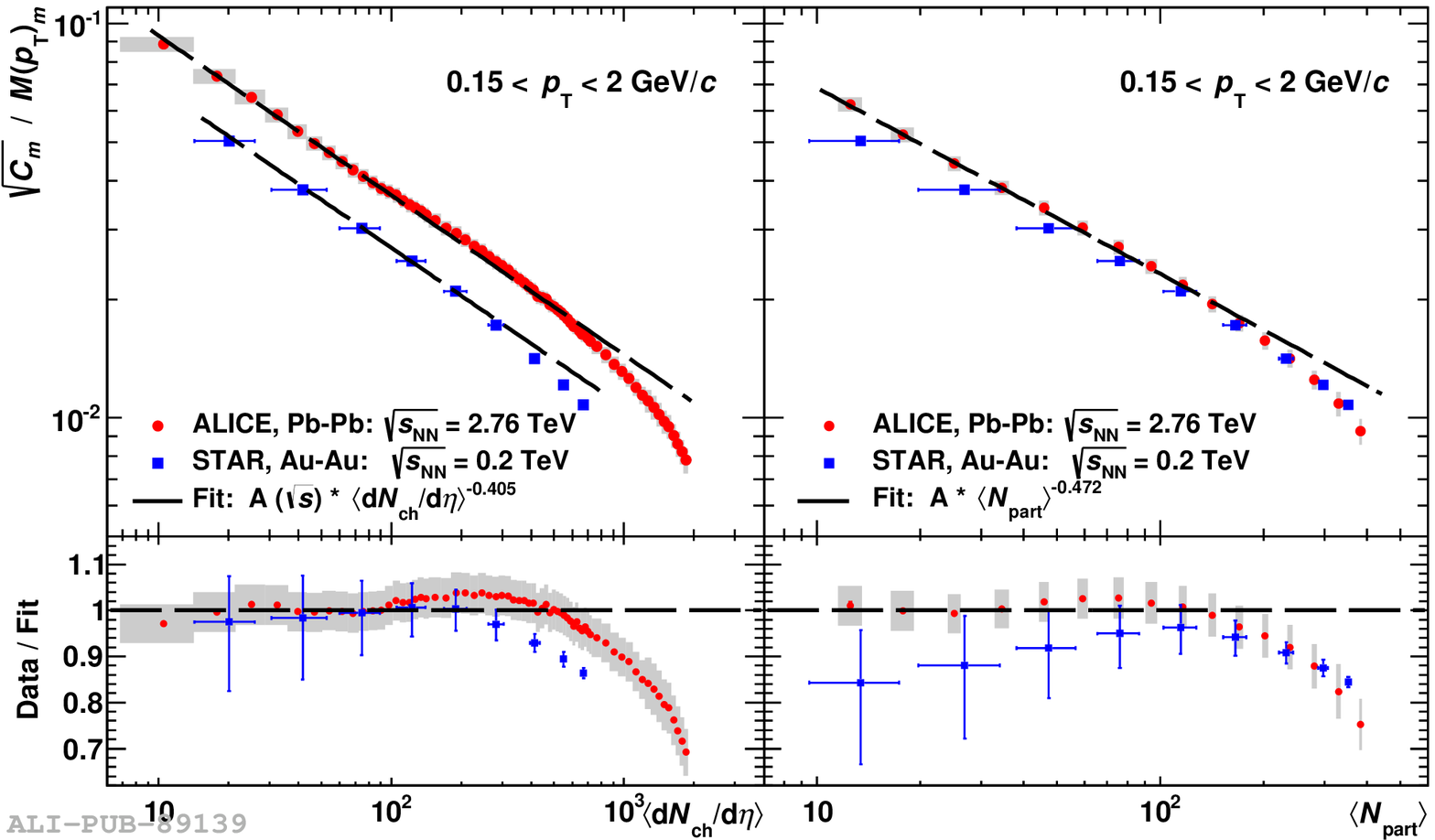}
(b)
\end{center}
\end{minipage} 
\caption{\label{pt} (a): Relative dynamical fluctuations $\sqrt{C_m}/{M(p_{\rm T})}_{m}$  as a function of $ \langle dN_{ch}/d\eta \rangle $ in pp and Pb$-$Pb collisions at $\sqrt{s_{NN}} =$ 2.76 TeV. Data are compared with the results from HIJING and power-law fitted to pp data and HIJING. (b): (left) Relative dynamical fluctuations $\sqrt{C_m}/{M(p_{\rm T})}_{m}$  as a function of $ \langle dN_{ch}/d\eta \rangle $ and (right) relative dynamical fluctuations $\sqrt{C_m}/{M(p_{\rm T})}_{m}$  as a function of $ \langle N_{part} \rangle $ for Pb$-$Pb collisions at $\sqrt{s_{NN}} =$ 2.76 TeV compared to STAR measurements in Au$-$Au collisions at $\sqrt{s_{NN}} =$ 0.2 TeV \cite{12}. Figures are taken from Ref. \cite{11}. }
\end{center}
\vskip -12pt
\end{figure}

By construction, $C_m$ vanishes for uncorrelated particle emission where only statistical fluctuations are present in the system. ALICE has measured the mean $p_{\rm T}$ fluctuations by taking a dimensionless ratio $\sqrt{C_m}/{M(p_{\rm T})}_{m}$, which quantifies the strength of the dynamical fluctuations in units of the average transverse momentum ${M(p_{\rm T})}_m$ in the multiplicity class $m$. The relative dynamical fluctuations $\sqrt{C_m}/{M(p_{\rm T})}_{m}$ as a function of average charged-particle multiplicity $ \langle dN_{ch}/d\eta \rangle $ in Pb$-$Pb collisions at $\sqrt{s_{NN}} =$ 2.76 TeV are shown in Figure \ref{pt} (a). The pp data are fitted with a power law, which is extrapolated to higher multiplicities. Peripheral Pb-Pb events are well in agreement with this pp
baseline. While moving from peripheral to most central collisions, a significant reduction of relative fluctuations is observed. In Figure \ref{pt} (b) (left) and Figure \ref{pt} (b) (right), comparisons of ALICE results of $\sqrt{C_m}/{M(p_{\rm T})}_{m}$ with Au$-$Au collision at $\sqrt{s_{NN}} =$ 0.2 TeV data measured by the STAR experiment are shown as a function of $ \langle dN_{ch}/d\eta \rangle $ and $ \langle N_{part} \rangle $, respectively.  The relative fluctuations for both energies are described by the pp baseline fit from peripheral up to mid-central collisions. However, data from both experiments deviate from the fit at the same centrality.  

%-------------------------------------------------
%-------------------------------------------------

%\begin{figure}[!h]
%\begin{center}$
%\begin{array}{cc}
%\includegraphics[width=2.15in]{2014-Sep-19-5_PtFluc_pp-PbPb-MC_relative_dNdeta.eps} &
%\includegraphics[width=3.4in]{2014-Sep-19-7_PtFluc_PbPb_relative_dNdeta_Npart.eps}
%\end{array}$
%\end{center}  
%\caption{Left: Relative dynamical fluctuations $\sqrt{C_m}/{M(p_{\rm T})}_{m}$  as a function of $ \langle dN_{ch}/d\eta \rangle $ in pp and Pb$-$Pb collisions at $\sqrt{s_{NN}} =$ 2.76 TeV. Data are compared with the results from HIJING and power-law fitted to pp data and HIJING. Middle and right: Relative dynamical fluctuations $\sqrt{C_m}/{M(p_{\rm T})}_{m}$  as a function of $ \langle dN_{ch}/d\eta \rangle $ and relative dynamical fluctuations $\sqrt{C_m}/{M(p_{\rm T})}_{m}$  as a function of $ \langle N_{part} \rangle $ for Pb$-$Pb collisions at $\sqrt{s_{NN}} =$ 2.76 TeV compared to STAR measurements in Au$-$Au collisions at $\sqrt{s_{NN}} =$ 0.2 TeV \cite{12}. Figures are taken from Ref. \cite{11}.}
%\label{pt}
%\end{figure}

\section{Summary and outlook}
In summary, the measurements of dynamical net-charge fluctuations are presented as a function of centrality and compared with theoretical expectations for HRG and QGP. That data are compared with previously measured results by different experiments at different collision energies. Moreover, the $\nu_{(+-,dyn)}^{corr}$  are measured as a function of the pseudorapidity windows.  The measurement of the width of balance function is found to decrease from peripheral to central collisions. We also presented the mean $p_{\rm T}$  fluctuation results for pp and Pb$-$Pb collisions. A decrease of the fluctuations with increasing multiplicity is observed in pp collisions. The mean $p_{\rm T}$  fluctuation results  for Pb$-$Pb and Au$-$Au collision data at $\sqrt{s_{NN}} =$ 2.76 TeV and 0.2 TeV deviate from a power-law fit to the pp data at the same centrality.  In future,  ALICE will extend its fluctuation studies to other analyses, like, multiplicity fluctuations, higher moments and cumulants of conserved  charge fluctuations. ALICE has the plan to extend the balance function studies to pp and p$-$Pb collisions both for charged particles and identified particles.

\end{document}